\documentclass[11pt,sort&compress]{elsarticle}
\usepackage[paper=letterpaper,top=24mm, bottom=26mm, left=26mm, right=26mm]{geometry}
\makeatletter
\def\ps@pprintTitle{%
 \let\@oddhead\@empty
 \let\@evenhead\@empty
 \def\@oddfoot{\centerline{\thepage}}%
 \let\@evenfoot\@oddfoot}
\makeatother
\usepackage{amsmath}
\usepackage{accents}
\usepackage{amssymb}
\usepackage{mathrsfs}
\usepackage{mathtools}
\usepackage[LGR,T1]{fontenc}
\usepackage[latin1]{inputenc}
\let\oldbibliography\thebibliography
\renewcommand{\thebibliography}[1]{%
  \oldbibliography{#1}%
  \setlength{\itemsep}{1.4pt}%
}
\usepackage[cal=boondox,calscaled=1]{mathalfa}
\DeclareMathAlphabet{\bbvar}{U}{BOONDOX-ds}{m}{n}
\DeclareMathAlphabet{\bbgreek}{U}{bbold}{m}{n}
\usepackage{graphicx}
\usepackage{psfrag}
\usepackage{pstool}
\usepackage{caption}
\usepackage{showlabels}
\usepackage{graphicx}%
\usepackage[colorlinks=true,
            linkcolor=blue,
            urlcolor=blue,
            citecolor=blue]{hyperref}
\usepackage[hyperref]{xcolor}
\definecolor{darkred}{rgb}{.95,.0,.0}

\newcommand{\qq}[1]{``#1''} 
\newcommand{\q}[1]{`#1'\,}  
\newcommand{\utilde}[1]{\underaccent{\tilde}{#1}}
\newcommand{\di}{\mathrm{d}}
\usepackage{tensor}\newcommand{\ou}[3]{{#1}{}^{#2}{}_{#3}}
\newcommand{\uo}[3]{{#1}{}_{#2}{}^{#3}}

\newcommand{\I}{\mathrm{i}} 
\newcommand{\E}{\mathrm{e}} 
\newcommand{\C}{\mathbb{C}}

\newcommand{\R}{\mathbb{R}}
\newcommand{\Z}{\mathbb{Z}}

\newcommand{\0}{o}

\newenvironment{subalign}{\subequations\align}{\endalign\endsubequations}
\newcommand{\eref}[1]{(\ref{#1})}

\newcommand{\AdS}{\mathrm{AdS}_{2}}

\newcommand\vpm{\mathbin{\vcenter{\hbox{
  \oalign{\hfil$\scriptstyle+$\hfil\cr
          \noalign{\kern-.3ex}
          $\scriptscriptstyle({-})$\cr}}}}}
\DeclareMathAlphabet{\sfit}{OT1}{fos}{sb}{it}
\DeclareMathAlphabet{\mathsf}{OT1}{fos}{sb}{n}

\definecolor{darkgreen}{rgb}{0.01, 0.75, 0.24}
\definecolor{darkblue}{rgb}{0.01, 0.24, 0.75}

\usepackage[multiple, flushmargin]{footmisc}

\let\originalleft\left
\let\originalright\right
\renewcommand{\left}{\mathopen{}\mathclose\bgroup\originalleft}
\renewcommand{\right}{\aftergroup\egroup\originalright}
\newcommand{\Tr}{\operatorname{Tr}}

\begin{document}

\begin{abstract}
\noindent Jackiw\,--\,Teitelboim (JT) gravity is a $1$+$1$-dimensional toy model for quantum gravity in four spacetime dimensions. 
In the absence of matter, JT gravity is a topological field theory and there are no local observables. The introduction of a boundary changes the situation. What was an unphysical gauge direction before turns into a physical boundary degree of freedom (a gravitational edge mode).  Starting from the $BF$ formulation of JT gravity, we develop a twistor representation for the boundary charges of JT gravity. We introduce point sources in the bulk and study the coupled gravity plus matter system at the quantum level. Eigenstates of quasi-local energy are built from the tensor product of unitary irreducible representations of $SL(2,\R)$. Empty patches of $\AdS$ are characterised by the continuous series representations, point sources are related to the discrete series.  Physical states are constructed by fusing the $SL(2,\R)$ representations into an $SL(2,\R)$ invariant singlet. The singlet lies in the kernel of the constraints, which are the analogue of the Wheeler\,--\,De Witt equations in the presence of distributional matter sources.\end{abstract}%

\title{Twistor representation of Jackiw\,--\,Teitelboim gravity}
\author{Wolfgang Wieland}
\address{Perimeter Institute for Theoretical Physics\\31 Caroline Street North\\ Waterloo, ON N2L\,2Y5, Canada\\{\vspace{0.5em}\normalfont 30 March 2020}
}
\maketitle
\vspace{-1.2em}
\hypersetup{
  linkcolor=black,
  urlcolor=black,
  citecolor=black
}
{\tableofcontents}
\hypersetup{
  linkcolor=black,
  urlcolor=darkred,
  citecolor=darkred,
}
\begin{center}{\noindent\rule{\linewidth}{0.4pt}}\end{center}\newpage
\section{Outline and motivation}
\noindent Jackiw\,--\,Teitelboim (JT) gravity \cite{Jackiw:1984je,Mann:1989gh,Teitelboim:1983ux,Nojiri:2000ja,Grumiller:2002nm,Grumiller:2017qao,Blommaert:2018iqz,Blommaert:2018oro,Iliesiu:2019xuh,Cangemi:1992bj,Isler:1989hq,Chamseddine:1989yz} is a toy model for quantum gravity in four spacetime dimensions. Spacetime is cut down to the $r$-$t$ plane and the transversal $\vartheta$-$\varphi$ directions are removed from the manifold. The only remnant of the transversal geometry is the dilaton $\Phi(r,t)$, which describes the area of the $r=\mathrm{const}.$ $t=\mathrm{const}.$ surfaces prior to the reduction to $1$+$1$ dimensions. Although JT gravity is topological, there are still important lessons to be learnt for quantum gravity in four dimensions. This has to do with gravitational edge modes\footnote{Such edge modes are well-known from Yang\,--\,Mills theory, where the introduction of a boundary turns otherwise unphysical (Coulombic) modes into physical boundary modes  \cite{Casini:2013rba,Donnelly:2015hxa,Gomes:2019xto,Gomes:2019rgg,Rovelli:2013fga,Donnelly:2016auv}. } on manifolds with boundaries \cite{Balachandran:1994up,Donnelly:2016auv,Freidel:2015gpa,Geiller:2017whh,Geiller:2019bti,Dittrich:2018xuk,Freidel:2019ees,Freidel:2019ofr,Wieland:2017zkf,Wieland:2018ymr,Asante:2018kfo}. The  characteristic initial value problem \cite{Bondi21,Sachs103,Sachs:1962zzb, Chrusciel:2012ap,Ashtekar:1978zz,AshtekarNullInfinity} provides a simple framework to understand the emergence of such edge degrees of freedom in gravity. At null infinity, initial data is characterised by the two radiative modes, namely the leading $O(r^{-2})$ term of the outgoing shear. Yet, the radiative data by itself is not enough to integrate the Einstein equations. We also need to specify corner data on a two-dimensional cross section of future (past) null infinity.  The question is then whether the corner data are physical or can be removed by appropriate gauge fixing conditions. 
The question is settled by the Hamiltonian analysis. On the covariant phase space \cite{AshtekarNullInfinity,Ashtekar:1990gc,Lee:1990nz,Wald:1999wa}, the pre-symplectic two-form has degenerate null directions. The gauge orbits lie tangential to these null directions, which determine the unphysical gauge directions on phase space. There is no gauge fixing that would remove the corner data completely, and JT gravity is a very simple framework to understand this mechanism in a simplified context. In fact, the only physical degrees of freedom of JT gravity are the edge modes alone. JT gravity is, therefore, a simple toy model to test the quantisation of such edge modes and compare the results with what we know in four dimensions. The physical relevance of such boundary modes has to do with the first law of black hole mechanics. For given energy and angular momentum, there is an exponentially large number of  boundary excitations at the horizon. The entropy is the logarithm of this number, which is proportional to the area of the horizon \cite{Strominger:1997eq,PhysRevD.51.632,Bianchi:2012ev,bhentashkrasn,Engle:2010kt,FGPfirstlaw,Ghosh:2011fc}.

The paper is organised as follows. \hyperref[sec2]{Section 2} provides a very brief introduction to the geometry of $\AdS$ and its symmetry group, which is $SL(2,\R)$. The goal of this introductory \hyperref[sec2]{section} is to wrap up the basic notation and clarify the formalism. \hyperref[sec3]{Section 3} is an important cornerstone for the rest of the paper. We introduce a spinor representation for the hyperbolic and elliptic elements of $\mathfrak{sl}(2,\R)$ and parametrise the elements of the Lie algebra in terms of their eigen spinors. Hyperbolic elements have two linearly independent and real-valued eigen spinors, an elliptic element, on the other hand, is characterised by one complex two-component eigen spinor. The parabolic case is the limiting case, where the complex spinor is real. These $\mathfrak{sl}(2,\R)$ eigen spinors are equipped with a natural $SL(2,\R)$  invariant Poisson structure. Upon quantising the Poisson structure, we recover the discrete and unitary series of the unitary representations of $SL(2,\R)$. In non-perturbative quantum gravity, such spinor representations characterise gravitational boundary modes at an entangling surface (such as a black hole horizon). Our construction provides, in fact, the $1$+$1$-dimensional analogue of the spinor-valued boundary modes \cite{twist,Livinerep,Bianchi:2016hmk} that have been studied recently in the context of non-perturbative quantum gravity \cite{Namburi:2019qja,Wieland:2017zkf,Wieland:2018ymr,wieland:nulldefects,Wieland:2019hkz,Wieland:2017cmf}. This analogy is laid out and developed in \hyperref[sec4]{section 4} and \hyperref[sec5]{section 5} of the paper. \hyperref[sec4]{Section 4} develops the classical bulk plus boundary theory. Hyperbolic elements of $\mathfrak{sl}(2,\R)$ describe quantum states of black hole spacetimes, elliptic elements, on the other hand, arise from spinning point particles in $\AdS$. The underlying gravitational action is the sum of the topological $\mathrm{Tr}(BF)$ action for JT gravity \cite{Isler:1989hq,Cangemi:1992bj,Chamseddine:1989yz} and a boundary term. The boundary term is added to impose the boundary conditions without violating the $SL(2,\R)$ gauge symmetry in the bulk. This is possible only at the expense of introducing additional boundary fields \cite{Donnelly:2016auv,Wieland:2017zkf}. The evolution of these boundary modes is governed by a boundary Hamiltonian. We derive the most general such Hamiltonian and study the gauge symmetries of the coupled bulk plus boundary system. The gauge symmetries are simultaneous $SL(2,\R)$ transformations of the bulk plus boundary fields and diffeomorphisms of compact support in the interior of the manifold. Finite diffeomorphisms that do not vanish at the boundary are physical symmetries generated by the boundary Hamiltonian. The gauge symmetries restrict the functional form of the  Hamiltonian to an arbitrary function of the $SL(2,\R)$ Casimir. In fact, the simplest boundary Hamiltonian is the $SL(2,\R)$ Casimir \cite{Cangemi:1992bj,Iliesiu:2019xuh,Maldacena:2016upp} itself. Finally, we study the theory at the quantum level. In the absence of matter and for fixed topology, the quantisation is trivial. Physical states are $SL(2,\R)$ invariant functions of the gravitational edge modes $q_\pm^A\in\R^2$ that are excited at either end of the strip.\footnote{The manifold has the topology of an infinite strip $[-1,1]\times\R$.}  If we add distributional point sources, the situation becomes more interesting. The gravity plus matter system is characterised by the fusion of unitary irreducible representation of $SL(2,\R)$. A matter defect carries a discrete series representation, the adjacent patches of empty $\AdS$ are characterised by the unitary continuous (principal) series representations. The constraints fuse these representation into an $SL(2,\R)$ invariant singlet (\hyperref[sec5]{section 5}). Finally, we conclude with a summary and discussion and explain how the results of this paper provide insights into various approaches to quantum gravity.

\section[Dyads and frame fields for AdS2]{Dyads and frame fields for $\boldsymbol{\AdS}$}\label{sec2}
\noindent In this section, we collect some very basic facts about the geometry of the spin bundle in two spacetime dimensions. The starting point is the introduction of a null co-frame $(k_a,\ell_a)$ that diagonalises the signature $(-,+)$ metric,
\begin{equation}
g_{ab} = - 2 k_{(a}\ell_{b)}.
\end{equation}
The corresponding (future pointing) vector fields $(\ell^a,k^a)$ are null and normalised to $k_a\ell^a=-1$. Internal Lorentz transformations act as dilations
\begin{subalign}
\tilde{k}_a& =\E^{+\lambda} k_a,\\
\tilde{\ell}_a &= \E^{-\lambda} \ell_a.
\end{subalign}
We say, therefore, that $k_a$ (resp.\ $\ell_a$) has Lorentz weight plus (minus) one, and we write, accordingly
\begin{equation}
k_a=\ou{e}{+}{a},\qquad \ell_a =\ou{e}{-}{a}.
\end{equation}
In addition to the frame fields, we also need a covariant derivative $\nabla = \di + \Gamma$. The corresponding spin connection $\Gamma_a$, which takes values in the Lie algebra of the Lorentz group $\R^\times =\R-\{0\}$, is uniquely determined by the torsionless equation,
\begin{equation}
T^\pm=\nabla e^\pm \equiv \di e^\pm \pm \Gamma\wedge e^\pm.\label{Tdef}
\end{equation}
The connection is abelian and the curvature tensor is simply given by
\begin{equation}
F=\di\,\Gamma \Leftrightarrow F_{ab}=2\partial_{[a}\Gamma_{b]}.
\end{equation}
To explain how $\nabla$ acts onto arbitrary tensor fields, we introduce  tensor components with respect to the dual null frame, and write for e.g.\ a vector field $V^a$,\begin{equation}
V^a = \frac{1}{\sqrt{2}}\big(V^-k^a+V^+\ell^a\big),\qquad \tilde{V}^\pm = \E^{\pm\lambda}V^\pm.
\end{equation}
The component functions $V^\pm$ have Lorentz charge plus (minus) one, and the torsionless derivative acts, therefore, via
\begin{equation}
\nabla_aV^\pm = \partial_a V^\pm \pm \Gamma_a V^\pm.
\end{equation}
This definition extends naturally to arbitrary Lorentz weights (tensors). The commutator of two such covariant derivatives defines the Riemann curvature tensor
\begin{equation}
\ou{R}{c}{dab}= \ou{\varepsilon}{c}{d}F_{ab},
\end{equation}
where we introduced the two-dimensional volume two-form
\begin{equation}
\varepsilon_{ab} = 2k_{[a}\ell_{b]}.
\end{equation}
In two dimensions, a two-form is dual to a scalar, and  $F_{ab}$ is therefore completely determined by the curvature scalar,
\begin{equation}
F_{ab}= -\frac{1}{2}\varepsilon_{ab}R,\qquad R= g^{ab}\ou{R}{c}{acb}.
\end{equation}
A simple example for such a geometry is provided by empty $\AdS$. The metric is given by
\begin{equation}
ds^2 = -\ell^2\big(\cosh^2 \eta\,\di \tau^2-\di\eta^2\big),\label{gdef}
\end{equation}
where $\Lambda=-1/\ell^2$ is the cosmological constant. The time coordinate is $\tau\in\R$ and $\eta\in\R$ is the spatial coordinate. A possible choice for a null co-frame for this metric is given by
\begin{equation}
e^\pm =-\frac{\ell}{\sqrt{2}}\big(\cosh\eta\,\di\tau \mp\di\eta\big).\label{dyad}
\end{equation}
The torsionless condition \eref{Tdef} determines the abelian spin connection $\Gamma$ uniquely. From equation \eref{dyad}, we immediately find, in fact
\begin{equation}
\Gamma = -\sinh\eta\,\di \tau.
\end{equation}
The field strength satisfies 
\begin{equation}
F =\di\,\Gamma = \frac{1}{\ell^2} e^+\wedge e^-.\label{Fdef}
\end{equation}

The underlying symmetry group is $SL(2,\R)$, which is the symmetry group of empty $\AdS$. The symmetry becomes manifest by introducing the  $SL(2,\R)$ generators
\begin{equation}
\tau_0=\frac{1}{2}\begin{pmatrix}1&0\\0&-1\end{pmatrix},\quad
\tau_-=\frac{1}{\sqrt{2}}\begin{pmatrix}0&0\\-1&0\end{pmatrix},\quad
\tau_+=\frac{1}{\sqrt{2}}\begin{pmatrix}0&+1\\0&0\end{pmatrix},\quad\label{taudef}
\end{equation}
and defining the non-abelian $SL(2,\R)$ connection
\begin{equation}
A =\tau_0 \Gamma +\frac{1}{\ell}\big(\tau_+e^++\tau_-e^-\big).\label{conndef}
\end{equation}
The torsionless equation \eref{Tdef} and the field equation \eref{Fdef} for the  curvature can be now rearranged into the single $SL(2,\R)$ flatness constraint
\begin{equation}
\ou{F}{A}{B}[A] = \di\ou{A}{A}{B} + \ou{A}{A}{C}\wedge \ou{A}{C}{B} =0,\label{Flatnss}
\end{equation}
where the spinor indices $A,B,C,\dots\in\{+\tfrac{1}{2},-\tfrac{1}{2}\}$ transform according to the matrix representation that  is implicitly defined by equation \eref{taudef}. 
\section[Spinor representation of {SL(2,R)}]{Spinor representation of $\boldsymbol{SL(2,\R)}$}\label{sec3}
\noindent The Poisson algebra of the gravitational edge modes for JT gravity is based on  the representation theory of $SL(2,\R)$, which becomes particularly transparent in terms of twistor-type variables \cite{penroserindler}. The representation is characterised by a non-degenerate\footnote{i.e.\ $\epsilon_{AB}\xi^B=0\Leftrightarrow\xi^A=0$.}  two-form $\epsilon_{AB}$ on $\R^2$. A possible representative  is given by \begin{equation}
\epsilon_{AB}\xi^A\eta^B=\xi^{+/2}\eta^{-/2}-\xi^{-/2}\eta^{+/2},\qquad \xi^A,\eta^A\in\R^2,
\end{equation}
where the spin up and down components have Lorentz weight plus (minus) one-half,
\begin{equation}
\nabla_a\xi^{\pm/2}=\partial_a\xi^{\pm/2}\pm\frac{1}{2}\Gamma_a\xi^{\pm/2}.
\end{equation}
The symmetry group of $\epsilon_{AB}=-\epsilon_{BA}$ can be now identified with $SL(2,\R)$ via
\begin{equation}
\Lambda\in SL(2,\R)\Leftrightarrow \Lambda\in GL(2,\R):\epsilon_{CD}\ou{\Lambda}{C}{A}\ou{\Lambda}{D}{B}=\epsilon_{AB}.\label{SLdef}
\end{equation}
Since the two-form $\epsilon_{AB}$ is non-degenerate, we can raise and lower indices,
\begin{equation}
\xi_A=\xi^B\epsilon_{BA},\qquad\xi^A=\epsilon^{AB}\xi_B,
\end{equation}
where $\epsilon^{AB}$ is the inverse two-form (a bivector in $\R^2$) such that $\epsilon^{AC}\epsilon_{BC}=\uo{\epsilon}{A}{B}=\delta^B_A$. Evaluating \eref{SLdef} in the vicinity of the unit element, we are now able to identify the Lie algebra $\mathfrak{sl}(2,\R)$ with those elements in $\ou{\lambda}{A}{B}\in\R^2\otimes(\R^2)^\ast$ that satisfy
\begin{equation}
\lambda_{AB}-\lambda_{BA}=0,
\end{equation}
hence $\lambda_{AB}=\lambda_{(AB)}=\epsilon_{CA}\ou{\lambda}{C}{B}$ is symmetric. Over the field of complex numbers, any symmetric rank two spinor $\omega_{AB}=\omega_{BA}\in(\C^2)^\ast\otimes (\C^2)^\ast$ has two complex eigen  spinors $\psi_A\in (\C^2)^\ast$ and $\phi_A\in (\C^2)^\ast$, such that the symmetrised tensor product satisfies 
\begin{equation}
\omega_{AB}=\psi_{(A}\phi_{B)},
\end{equation}
see \cite{penroserindler} for a simple proof based on the fundamental theorem of algebra. For any given $\omega_{AB}\neq 0$, the eigen spinors $(\psi_A,\phi_A)$ are uniquely determined modulo (i) complex rescalings $(\psi_A,\phi_A)\rightarrow(\E^{+z}\psi_A,\E^{-z}\phi_A)$ with $z\in\C$ and (ii) the exchange $(\psi_A,\phi_A)\rightarrow(\phi_A,\psi_A)$ of the two elements. In terms of the real spinor representation of $SL(2,\R)$, which we used in equation \eref{SLdef} above, the $SL(2,\R)$ Lie algebra is identified with traceless $2\times 2$ matrices $\ou{\lambda}{A}{B}$, whose entries are real $\ou{\lambda}{A}{B}\in\R^2\otimes(\R^2)^\ast\subset\C^2\otimes(\C^2)^\ast$. The  symmetrised tensor product must satisfy, therefore, the following additional reality conditions,
\begin{equation}
\lambda_{AB}=\psi_{(A}\phi_{B)}=\bar{\psi}_{(A}\bar{\phi}_{B)}=\bar{\lambda}_{AB}.
\end{equation}
Now, for any given $\lambda_{AB}=\lambda_{BA}\neq 0$, the eigen spinors $\psi_A$ and $\phi_B$ are unique only up to the exchange $(\psi,\phi)\rightarrow(\phi,\psi)$ and the complex-valued rescalings $(\psi,\phi)\rightarrow(\E^z\psi,\E^{-z}\phi)$. There are therefore only two possibilities for how the ordered pair $(\psi,\phi)$ is related to $(\bar\psi,\bar\phi)$, and we classify the two possibilities as follows,
\begin{subalign}
\text{\emph{continuous series:}}&\quad\psi_A=\E^z \bar{\psi}_A,\qquad \phi_A=\E^{-z}\bar{\phi}_A,\\
\text{\emph{discrete series:}}&\quad\psi_A=\E^z \bar{\phi}_A,\qquad \phi_A=\E^{-z}\bar{\psi}_A,
\end{subalign}
for some $z\in\C$. In the first case, we have $\psi_A=\E^z \bar{\psi}_A= \E^{z+\bar{z}}{\psi}_A$. Thus, $\E^{z}=\E^{\I\varphi}$ is a phase, for some $\varphi\in[0,2\pi)$. The rescaled spinors $P_A = \E^{-\I\frac{\varphi}{2}}\psi_A$ and $Q_A=\E^{\I\frac{\varphi}{2}}\phi_A$ are therefore real, and we find the parametrisation
\begin{equation}
\text{\emph{continuous series:}}\quad \lambda_{AB}=P_{(A}Q_{B)},\qquad P^A, Q^A\in\R^2,
\end{equation}
where $P_A$ and $Q_A$ are now unique only up to the exchange map $(P_A,Q_A)\rightarrow (Q_A,P_A)$ and the real-valued dilations $(P_A,Q_A)\rightarrow(\E^{+r}P_A,\E^{-r}Q_A)$ for some $r\in\R$. In the second case, we have $\psi_A=\E^z \bar{\phi}_A=\E^{z-\bar{z}}{\psi}_A$. Thus,  $\E^z=\pm\E^{r}$ is real-valued for some $r\in\R$. Replacing $\psi_A$ by the rescaled spinor $\E^{-\frac{r}{2}}\psi_A$, we obtain, therefore, the following parametrisation for the discrete series  elements,
\begin{equation}
\text{\emph{discrete series:}}\quad \lambda_{AB}=\pm\psi_{(A}\bar{\psi}_{B)},\qquad \psi^A\in\C^2,
\end{equation}
where $\psi_A$ is now unique only up to (i) the $U(1)$ transformation $\psi_A\rightarrow \E^{\I\varphi}\psi_A$ for some $\varphi\in[0,2\pi)$ and (ii) the complex conjugation exchanging $\psi_A$ with $\bar{\psi}_A$. The two series correspond to the discrete and continuous series representations of $SL(2,\R)$ respectively. This becomes obvious by equipping the spinors with the natural $SL(2,\R)$ invariant Poisson brackets\footnote{All other Poisson brackets among the fundamental variables vanish. There could also be an arbitrary constant on the right hand side of \eref{brackt1} and \eref{brackt2}, which we reabsorbed back into the definition of $\epsilon_{AB}$.}
\begin{subalign}
\text{\emph{continuous  series:}}&\quad\{P_A,Q_B\}=\epsilon_{AB},\label{brackt1}\\
\text{\emph{discrete series:}}&\,\quad\{\bar{\psi}_A,\psi_{B}\}=\label{brackt2}\epsilon_{AB}.
\end{subalign}
The $SL(2,\R)$ Casimir operator is $c= \mathrm{Tr}(\lambda^2)=-\lambda_{AB}\lambda^{AB}$. In the continuous series representation, we have $c = \frac{1}{2}(P_AQ^A)^2\geq 0$. In terms of the Schrödinger position representation, the momentum spinor  $P_A$ turns into the derivative operator $P_A=-\I\partial/\partial Q^A$ on $L^2(\R^2,d^2Q)$. The Casimir operator can be expressed, therefore, in terms of the dilation operator $-\I (Q^A\partial/\partial Q^A+1)$, whose spectrum is \emph{continuous} and real. In the discrete series representation, on the other hand, the eigenvalues of the Casimir are discrete. This can be seen as follows. Consider the spin up and down components $p=\psi^{+/2}$ and $z=\bar{\psi}^{-/2}$ of $\psi^A$ that satisfy the Poisson commutation relations
\begin{equation}
\{p,z\}=\{\bar{p},\bar{z}\}=1.
\end{equation}
The Casimir operator is $c= \mathrm{Tr}(\lambda^2)=-\lambda_{AB}\lambda^{AB}=\frac{1}{2}(\epsilon_{AB}\bar{\psi}^A\psi^B)^2\leq 0$, and it can be written, therefore, as (minus) the square of the angular momentum generator $\frac{1}{2\I}\epsilon_{AB}\bar{\psi}^A\psi^B= (pz-\bar{p}\bar{z})/(2\I)$ on the complex plane, which has a discrete spectrum in the quantum theory.
\section{Heisenberg boundary charges in JT gravity}
\subsection{Field equations in the interior}\label{sec4}
\noindent The action for Jackiw\,--\,Teitelboim gravity in the interior of the manifold is given by the topological $SL(2,\R)$ $BF$-action \cite{Cangemi:1992bj,Iliesiu:2019xuh,Isler:1989hq,Chamseddine:1989yz}
\begin{equation}
S[B,A]=\int_{\mathcal{M}}\ou{B}{A}{B}\ou{F}{B}{A}[A]\equiv\int_{\mathcal{M}}\operatorname{Tr}(BF).\label{BFactn}
\end{equation}
At the saddle point of the action \eref{BFactn}, the flatness constraint \eref{Flatnss}  is satisfied and the $B$ field is covariantly constant with respect to the $SL(2,\R)$ connection:  $DB=\di B+[A,B]=0$. The relation to the metric formulation is provided by the decomposition
\begin{equation}
B = -\frac{1}{4\pi G} \Phi \tau_0 +2\ell \big(p^-\tau_-+p^+\tau_+\big),
\end{equation}
where $\Phi$ is the dilaton and $p^\pm$ are the translational moments. Since $B$ is covariantly constant with respect to the $SL(2,\R)$ covariant derivative, we obtain
\begin{subalign}
\di\Phi& = 8\pi G\big(p^+e^--p^-e^+\big),\label{DB1}\\
\nabla p^\pm & = \di p^\pm\pm \Gamma p^\pm= \mp\frac{1}{8\pi G\ell^2} \Phi\, e^\pm\label{DB2}.
\end{subalign}
Inserting \eref{DB2} back into \eref{DB1}, we obtain the second order equation
\begin{equation}
\nabla_a\nabla_b\Phi = \frac{1}{\ell^2}\Phi g_{ab}.\label{DDphi}
\end{equation}
Its general solution in the $\AdS$ background  \eref{gdef} is determined by three integration constants $a$, $b$, and $\tau_o$ that correspond to the three independent directions in $\mathfrak{sl}(2,\R)$,
\begin{equation}
\Phi(\tau,\eta) = a \cos(\tau-\tau_o) \cosh\eta + b\sinh \eta.\label{Phisol}
\end{equation}
Notice now that if both $\nabla_a\Phi$ and $\Phi$ vanish at a point in the manifold, the dilaton $\Phi$ will vanish everywhere. Since  the Hessian $\nabla\otimes\nabla\Phi$ of the dilaton is proportional to the metric, and since the metric has Lorentzian $(-$$+)$ signature, the dilaton cannot have a minimum (maximum) inside the manifold. For $\Phi\neq 0$, the gradient $\nabla_a\Phi$ can only vanish at a saddle point. Going back to \eref{Phisol} and computing the gradient of $\Phi$, we then see that there can only be one such saddle point; and a saddle point exists, if and only if $|b/a|< 1$, which determines the sign of the Casimir\begin{equation}
2c =2 \Tr(B^2) = \frac{1}{(4\pi G)^2}\Big(\Phi^2-\frac{1}{\ell^2}g^{ab}\nabla_a\Phi\nabla_b\Phi\Big) =  \frac{1}{(4\pi G)^2} (a^2 - b^2).
\end{equation}
In other words, the dilaton admits a saddle point if an only if the Casimir is positive. What is the physical meaning of this statement? Jackiw\,--\,Teitelboim gravity is a symmetry reduced toy model for gravity in four dimensions. The $r$-$t$ plane is kept dynamical, the transversal $\vartheta$-$\varphi$ directions are removed from the theory. The only remnant of the transversal geometry is the dilaton, which determines the transversal area element: $d^2v = \Phi\,\sin^2\vartheta \,\di\vartheta\, \di\varphi$. The expansion of the area element along the null generators is determined by the gradient of $\Phi$. At the bifurcation surface of a black hole horizon, the expansion vanishes along both null generators. Such a surface can be identified, therefore, with the saddle points of the dilaton,
\begin{equation}
\vartheta_{(\ell)} = \Phi^{-1}\ell^a\nabla_a\Phi = 0,\qquad\vartheta_{(k)}=\Phi^{-1}k^a\nabla_a\Phi = 0.
\end{equation}
Such a saddle point exists if and only if the $SL(2,\R)$ Casimir is strictly greater than zero. A black hole geometry is therefore modelled by selecting the continuous series representations, where the auxiliary $B$ field assumes the following parametrisation,
\begin{equation}
B^{AB}\big|_{\text{BH}} = P^{(A}Q^{B)},\qquad P^A, Q^A\in\R^2.
\end{equation}

The discrete series representations, on the other hand, arise from the coupling to point particles via the worldline action
\begin{equation}
S_\gamma[\psi,\bar\psi,N|A] = -\int_\gamma\Big[\bar\psi_A D\psi^A-\frac{N}{2}\big(\I \bar\psi_A\psi^A-s\big)\Big],\label{partcl}
\end{equation}
 for a point particle whose mass $M$ and boost angular momentum $K$ are constrained to satisfy $s^2= 4(\ell^2M^2-K^2)\geq 0$ for some constant $s$. The lapse function $N$ imposes the corersponding constraint on $s$. Mass $M$ and spin $K$ are functions of the spin up and down components $\psi^{\pm/2}$ of the $SL(2,\R)$ spinor $\psi^A$. In fact, by reintroducing the $\psi^{\pm/2}$ components  of $\psi^A$, and spliting the $SL(2,\R)$ connection into its boost and translation components, we find
\begin{align}\nonumber
S_\gamma = \int_\gamma\Big[p^- e^++p^+ e^-+\Gamma K+\bar\psi^{-/2}\di\psi^{+/2}-\bar\psi^{+/2}&\di\psi^{-/2}+\\&+N\Big(\operatorname{\mathfrak{Im}}(\bar\psi^{-/2}\psi^{+/2})-\frac{s}{2}\Big)\Big].
\end{align}
The first two terms are the kinetic term for a particle propagating in a geometry determined by the metric $g_{ab}=-2\ou{e}{+}{(a}\ou{e}{-}{b)}$. The components of the particles momentum vector are given by
\begin{equation}
p^\pm = -p^a\ou{e}{\pm}{a} =  \frac{1}{\ell}\frac{1}{\sqrt{2}} \big|\psi^{\pm/2}\big|^2\geq 0. 
\end{equation}
The mass of the particle is $g_{ab}p^ap^b =-2p^+p^-= -M^2$. The boost angular momentum, on the other hand, can be inferred from the coupling to the spin connection $\Gamma$, and is  given by $K = \tfrac{1}{2}(\bar{\psi}^{-/2}\psi^{+/2}+\bar{\psi}^{+/2}\psi^{-/2})$.

\subsection{Bulk plus boundary action}
\noindent Depending on the  boundary conditions chosen, different boundary terms need to be added to the action. In the following, we will impose a specific class of Dirichlet boundary conditions, where {the boundary value of the spin connection $\Gamma_a$ and the frame field $\ou{e}{\pm}{a}$ are unconstrained. In addition, we restrict ourselves to a manifold, which has the topology of an infinite strip, i.e.\  $\mathcal M\simeq [-1,1]\times\R=\text{\q{\emph{space\,$\times$\,time}}}$.  For the continuous series representations that correspond to black holes spacetimes \cite{Jackiw:1984je,Mann:1989gh,Teitelboim:1983ux,Grumiller:2002nm}, the  bulk plus boundary action for such extended boundary conditions \cite{Donnelly:2016auv,Wieland:2017zkf} is given by
\begin{equation}
S[B,A|P,Q]=\int_{\mathcal{M}}\ou{B}{A}{B}\ou{F}{B}{A}[A]\mp\sum_\pm \int_{\gamma^\pm}\Big(P_ADQ^A -\kappa\,H_\pm[P,Q]\Big),\label{BFactn}
\end{equation}
where $\kappa$ is a one-form along the boundary, $D=\di+[A,\cdot]$ is the $SL(2,\R)$ covariant derivative and $H[P,Q]$ is a yet unspecified boundary Hamiltonian. We will identify the most general such Hamiltonian below. The boundary $\partial\mathcal{M}=\gamma_-^{-1}\cup\gamma_+$ has two disconnected parts, which have the same orientation.\footnote{Our conventions for the orientation of $\mathcal{M}$ are determined by $\int_{ \mathcal{M}}\di \omega =\int_{\gamma_+}\omega-\int_{\gamma_-}\omega$ for all one-forms $\omega$ in $\mathcal{M}$.} The one-form $\kappa$ is a $c$-number (a background field).  All other configuration variables are dynamical: a point in the infinite-dimensional space of histories is parametrised by a quadruple of bulk plus boundary fields $(B,A|P,Q)$.
 For the discrete series representations, the analogous bulk plus boundary action is given by
\begin{equation}
S[B,A|\psi,\bar{\psi}]=\int_{\mathcal{M}}\ou{B}{A}{B}\ou{F}{B}{A}[A]\mp \sum_\pm\int_{\gamma^\pm}\Big(\bar{\psi}_AD\psi^A -\kappa\,H_\pm[\bar{\psi},\psi]\Big).\label{BFactn2}
\end{equation}
The equations of motion in the bulk are the constraints \eref{DB1} and \eref{DB2} on the auxiliary $B$ field and the flatness condition for the $SL(2,\R)$ curvature, i.e.\ \eref{Flatnss}. Having introduced auxiliary boundary fields, we obtain, however, also additional boundary field equations. First of all, there are the gluing conditions, which follow from the variation of the connection. The $SL(2,\R)$ connection  enters both the action in the bulk and the boundary term. Using $\delta F = D\delta A$, and performing a partial integration, we obtain
\begin{equation}
\delta_A S[B,A|P,Q]=\int_{\partial\mathcal{M}} (DB_{AB})\wedge\delta A^{AB}\pm\sum_\pm\int_{\gamma_\pm}\big(P_{(A}Q_{B)}-B_{AB}\big)\delta A^{AB},
\end{equation}
and equally for the discrete series representations. From the saddle points of the coupled bulk plus boundary action, we thus learn that $B_{AB}$ is covariantly constant in the interior and that its boundary value is determined by the following \emph{gluing conditions},
\begin{subalign}
\text{\emph{continuous series:}}&\quad B_{AB}\big|_{\gamma_\pm}=P_{(A}Q_{B)},\label{glucond1}\\
\text{\emph{discrete series:}}&\quad B_{AB}\big|_{\gamma_\pm}=\psi_{(A}\bar{\psi}_{B)}.\label{glucond2}
\end{subalign}
In addition to the gluing conditions, there are the boundary field equations, which are obtained  from the variation of the boundary spinors themselves. If $\xi^a$ denotes a  vector field,  tangential to the boundary, the boundary field equations are simply given by the Hamilton equations
\begin{align}
\xi^aD_aQ^A & = +\kappa_\xi \frac{\partial H}{\partial P_A},\label{EOMb1}\\
\xi^aD_aP_A & = -\kappa_\xi \frac{\partial H}{\partial Q^A},\label{EOMb2}\\
\xi^aD_a\psi^A & =+ \kappa_\xi \frac{\partial H}{\partial \bar{\psi}_A},\label{EOMb3}
\end{align}
where $\kappa_\xi = \kappa_a\xi^a$. Since $B_{AB}$ is covariantly constant in the interior, it must also be covariantly constant along the boundary. This is only possible if the Hamiltonian  Poisson commutes with $P_{(A}Q_{B)}$ (resp.\ $\psi_{(A}\bar{\psi}_{B)}$), which are the generators of $SL(2,\R)$ transformations. The Hamiltonian must be therefore an $SL(2,\R)$ invariant function of $P_A$ and $Q^A$ (resp.\ $\bar{\psi}_{A}$ and ${\psi}^{A}$). The only  $SL(2,\R)$ invariant polynomials of $P^A$ and $Q_A$ (resp.\ $\bar{\psi}_{A}$ and ${\psi}^{A}$) are $(P_AQ^A)^n$ (resp. $(\I{\bar{\psi}}_A\psi^A)^n$), and the Hamiltonian must be therefore a function of $P_AQ^A$ (resp. $\I\bar{\psi}_A\psi^A$) alone. In other words,
\begin{align}
H[P,Q] &= f\big(P_A Q^A\big),
\end{align}
for some function $f:\R\rightarrow\R$.


\subsection{Hamiltonian formalism, gauge symmetries, observables}\label{sec3.3}
\noindent In the last section, we already briefly skimmed the Hamiltonian formulation of the bulk plus boundary theory. Let us now complete the analysis. To introduce the phase space, we choose an arbitrary Cauchy hypersurface $\Sigma$ that is anchored at points $c_{\pm}=\gamma_\pm\cap\Sigma$ at the boundary. The one-dimensional manifold $\Sigma$ carries an orientation, our conventions are as follows,
\begin{equation}
\int_\Sigma \di f = f\big|_{c_-}- f\big|_{c_+}\equiv\mp\sum_\pm f,
\end{equation}
for all functions $f$. In the following, we will base our analysis on the covariant phase space approach \cite{AshtekarNullInfinity,Ashtekar:1990gc,Lee:1990nz,Wald:1999wa}. The covariant phase space $\mathcal{P}_\Sigma$ is the space of solutions of the bulk plus boundary field equations in a neighbourhood of $\Sigma$. If we split the strip $\mathcal{M}\simeq [-1,1]\times\R$ along $\Sigma$ into two halves, and compute the first variation of the action, we obtain the covariant pre-symplectic potential for either representation,
\begin{subalign}
\text{\emph{continuous series:}}&\quad \Theta_\Sigma = \int_\Sigma \ou{B}{A}{B}\bbvar{d}\ou{A}{B}{A} \mp\sum_\pm P_A\bbvar{d}Q^A,\label{symplctc1}\\
\text{\emph{discrete series:}}&\quad \Theta_\Sigma = \int_\Sigma \ou{B}{A}{B}\bbvar{d}\ou{A}{B}{A} \mp\sum_\pm \bar{\psi}_A\bbvar{d}\psi^A,\label{symplctc2}
\end{subalign}
where $\bbvar{d}$ denotes the exterior derivative on phase space. Poisson brackets among physical observables are inferred from the pre-symplectic two-form $\Omega_\Sigma=\bbvar{d}\Theta_\Sigma$, which also encodes the gauge symmetries of the theory and identifies them with the  null vectors $\delta_{\text{gauge}}:\Omega_\Sigma(\delta_{\text{gauge}},\cdot)=0$. Simultaneous and infinitesimal $SL(2,\R)$ transformations of the bulk plus boundary fields are the simplest such vector fields on phase space. The bulk fields transform according to the left action,
\begin{align}
\delta_\lambda \ou{B}{A}{B} & = \ou{\lambda}{A}{C}\ou{B}{C}{B}-\ou{B}{A}{C}\ou{\lambda}{C}{B},\label{g1}\\
 \delta_\lambda \ou{A}{A}{Ba} & = -D_a\ou{\lambda}{A}{B}.\label{g2}
\end{align}
where $\lambda\in\mathcal{C}^1(\mathcal{M}:\mathfrak{sl}(2,\R))$ is the gauge element. The boundary fields transform equally,
\begin{equation}\left.
\begin{split}
\delta_\lambda P_A &= -\ou{\lambda}{C}{A}P_C,\\
\delta_\lambda Q^A &= +\ou{\lambda}{A}{C}Q^C
\end{split}\qquad
\begin{split}
\delta_\lambda \bar\psi_A & = - \ou{\lambda}{C}{A}\bar{\psi}_C,\\
\delta_\lambda \psi^A & = +\ou{\lambda}{A}{C}\psi^C.\label{g3}
\end{split}\qquad\right\}
\end{equation}
A short calculation gives for e.g.\ the continuous series that
\begin{align}\nonumber
\Omega_\Sigma(\delta_\lambda,\delta) & = \int_\Sigma\Big[2\ou{\lambda}{A}{C}\ou{B}{C}{B}\delta\ou{A}{B}{A}+\delta[\ou{B}{B}{A}]D\ou{\lambda}{A}{B}\Big]\pm\sum_\pm\ou{\lambda}{A}{B}\delta[P_AQ^B]=\\\nonumber
&= \int_\Sigma\Big[-\ou{\lambda}{A}{B}[\delta,D]\ou{B}{B}{A}-\ou{\lambda}{A}{B}[D,\delta]\ou{B}{B}{A}\Big]\pm\sum_\pm\ou{\lambda}{A}{B}\delta\big[P_AQ^B-\ou{B}{B}{A}\big]=\\
&=\mp\sum_\pm\lambda^{AB}\delta\big[P_{(A}Q_{B)}-B_{AB}\big]=0.
\end{align}
Going from the first to the second line, we performed a partial integration and used the vanishing of the covariant differential $D\ou{B}{A}{B}$ on the covariant phase space, which implies,
\begin{equation}
2 \delta\ou{A}{C}{(A}B_{C)B}=[D,\delta]B_{AB}  = D[\delta B_{AB}].
\end{equation}
We thus see that the simultaneous bulk plus boundary $SL(2,\R)$ transformations (\ref{g1}, \ref{g2}, \ref{g3}) define indeed an unphysical gauge direction in $T\mathcal{P}_\Sigma$,
\begin{equation}
\Omega_\Sigma(\delta_\lambda,\cdot) = 0.\label{g4}
\end{equation}

In a similar way, let us now consider  the action of diffeomorphisms on the covariant phase space. Any diffeomorphism $\varphi$, which is smoothly connected to the identity, is generated by a vector field $\xi^a\in T\mathcal{M}$ via the exponential map $\varphi = \exp(\xi)$. The Lie derivative\footnote{The symbol \qq{$\di$} denots the exterior derivative and \qq{$\iota_\xi$} is the interior product.} $L_\xi[\cdot]=\di[\iota_\xi(\cdot)]+\iota_\xi[\di(\cdot)]$ lifts any such vector field into a vector field  on the covariant phase space, $L_\xi\in T\mathcal{P}_\Sigma$. The boundary spinors are fields that are intrinsic to the boundary. They can only be Lie dragged along the boundary, and we restrict ourselves therefore to only those bulk vector fields $\xi^a\in T\mathcal{M}$ that lie tangential to the boundary,
\begin{equation}
\xi^a\big|_{\gamma_\pm} \in T\gamma_\pm.
\end{equation}
Such vector fields are Hamiltonian, because there exists a Hamilton function $\boldsymbol{H}_\xi$ such that the Hamilton equations are satisfied, i.e.\
\begin{equation}
\delta[\boldsymbol{H}_\xi] = -\Omega_\Sigma(L_\xi,\delta),
\end{equation}
for all vector fields on phase space $\delta\in T\mathcal{P}_\Sigma$. All fundamental configuration variables are charged under $SL(2,\R)$, and it is therefore useful to introduce a gauge covariant extension of the ordinary Lie derivative $L_\xi$. This is achieved by adding an infinitesimal $SL(2,\R)$ gauge transformation that restores the gauge symmetry,
\begin{equation}
\mathcal{L}_\xi = L_\xi + \delta_{\lambda[\xi,A]},\qquad \ou{\lambda}{A}{B}[\xi,A] = \xi^a\ou{A}{A}{Ba}.
\end{equation}
We have seen in above, see \eref{g4}, that any infinitesimal $SL(2,\R)$ gauge element $\ou{\lambda}{A}{B}:\mathcal{M}\rightarrow\mathfrak{sl}(2,\R)$ defines a degenerate direction $\delta_\lambda\in T\mathcal{P}_\Sigma$ of the pre-symplectic two-form. It does not matter, therefore, whether we use the ordinary Lie derivative or the gauge covariant derivative to compute the Hamiltonian, 
\begin{equation}
\delta[\boldsymbol{H}_\xi]  =- \Omega_\Sigma(\mathcal{L}_\xi,\delta)=-\Omega_\Sigma(L_\xi,\delta).
\end{equation}
Given the infintitesimal $SL(2,\R)$ transformations $\delta_\lambda[\ou{A}{A}{Ba}]=-D_a\ou{\lambda}{A}{B}$, we obtain the gauge covariant Lie derivative of the connection,
\begin{align}\nonumber
\mathcal{L}_\xi \ou{A}{A}{Ba} &=L_\xi\ou{A}{A}{Ba} +\delta_{\lambda[\xi,A]}[\ou{A}{A}{Ba}]  = L_\xi \ou{A}{A}{Ba} - D_a(\xi^b\ou{A}{A}{Bb})=\\
&=\xi^b\partial_b\ou{A}{A}{B a}-\xi^b\partial_a\ou{A}{A}{Bb}+\xi^b\big(\ou{A}{A}{Cb}\ou{A}{C}{Ba}-\ou{A}{A}{Ca}\ou{A}{C}{Bb}\big)=\xi^b\ou{F}{A}{Bba},
\end{align}
where $\ou{F}{A}{Bab}$ is the curvature two-form. In the same way, we obtain the gauge covariant Lie derivative of the auxiliary $B$ field,
\begin{align}\nonumber
\mathcal{L}_\xi \ou{B}{A}{B} &= L_\xi\ou{B}{A}{B}+\delta_\lambda[\ou{B}{A}{B}]
=\xi^bD_b\ou{B}{B}{A}.
\end{align}
where $\lambda\in\mathcal{C}^1(\mathcal{M}:\mathfrak{sl}(2,\R))$ is the gauge element. The boundary fields transform equally,
\begin{equation}\left.
\begin{split}
\mathcal{L}_\xi P_A &= \xi^aD_aP_A,\\
\mathcal{L}_\xi Q^A &= \xi^aD_aQ^a,
\end{split}\qquad
\begin{split}
\mathcal{L}_\xi\bar\psi_A & = \xi^aD_a\bar{\psi}_A,\\
\mathcal{L}_\xi\psi^A & = \xi^aD_a\psi^A.\label{d3}
\end{split}\qquad\right\}
\end{equation}
Going back to the bulk and boundary field equations of motion, namely (\ref{Flatnss}, \ref{DB1}, \ref{DB2}), and (\ref{EOMb1}, \ref{EOMb2}) respectively, we thus find for e.g.\ the continuous series representations,
\begin{align}\nonumber
\Omega_\Sigma(\mathcal{L}_\xi,\delta) & = \int_\Sigma \Big(\mathcal{L}_\xi[\ou{B}{A}{B}]\delta[\ou{A}{B}{A}]-\delta[\ou{B}{A}{B}]\mathcal{L}_\xi[\ou{A}{B}{A}]\Big)\mp\sum_\pm\big(\mathcal{L}_\xi[P_A]\delta[Q^A]-\delta[P_A]\mathcal{L}_\xi[Q^A]\big)=\\
&=\pm\sum_\pm\kappa_\xi\left(\frac{\partial H_\pm}{\partial Q^A}\delta Q^A+\frac{\partial H_\pm}{\partial P_A}\delta P_A\right)=\pm \sum_\pm\kappa_\xi\delta[H_\pm].
\end{align}
We may therefore infer the Hamiltonian, which is simply the Hamiltonian (\ref{EOMb1}, \ref{EOMb2}) of the boundary modes alone, 
\begin{equation}
\boldsymbol{H}_\xi = \mp \sum_\pm\kappa_\xi^\pm H_\pm=\mp \sum_\pm\kappa_\xi^\pm\, f_\pm(P_AQ^A)\big|_{c_\pm},
\end{equation}
where $\kappa_\xi^\pm = \xi^a\kappa_a\big|_{c_\pm}$. An analogous result also holds for the discrete series representations, where the generator $\boldsymbol{H}_\xi$ of diffeomorphisms $\exp(\xi)$ is again the sum of the Hamiltonian $H_\pm[\psi,\bar{\psi}]$ of the boundary modes $\psi^A$, see \eref{BFactn2}.

\subsection{Physical phase space}
\noindent The symplectic structure on the physical phase space is obtained by the pull-back of the pre-symplectic potential to the solution space of the bulk plus boundary field equations. The general solution for a flat $SL(2,\R)$ connection in the interior of the manifold is given by
\begin{equation}
A = g^{-1}\di g,\label{Asolvd1}
\end{equation}
where $g:\mathcal{M}\rightarrow SL(2,\R)$ denotes a finite $SL(2,\R)$ gauge transformation. The exterior functional derivative of such a flat connection can be expressed in terms of the functional Maurer\,--\,Cartan form $\bbvar{d}g\,g^{-1}$,  
\begin{equation}
\bbvar{d}[A] = g^{-1}\di\big(\bbvar{d}gg^{-1}\big) g\label{Asolvd2}
\end{equation}
Next, we consider the $B$ field, which is the momentum conjugate to the $SL(2,\R)$ connection. On shell, the $B$ field is covariantly constant. The general solution of such a covariantly constant $B$ field can be written in terms of the gauge elements $g$ and a constant $\mathfrak{sl}(2,\R)$ element,
\begin{equation}
b: = g Bg^{-1},\qquad DB = 0\Leftrightarrow \di b=0.\label{Bsolvd}
\end{equation}
Next, we introduce the analogous parametrisation of the boundary modes,
\begin{align}
p^A &=\ou{g}{A}{B}P^B,\label{dressd1}\\
q^A &=\ou{g}{A}{B}Q^B.\label{dressd2}
\end{align}
If we  insert the parametrisation \eref{Asolvd2}, \eref{Bsolvd} back into the pre-symplectic potential, we obtain 
\begin{align}\nonumber
\Theta_\Sigma & = \int_\Sigma \Tr\left(B\bbvar{d}A\right)\mp\sum_\pm P_A\bbvar{d}Q^A=\\\nonumber
& =\int_{\Sigma}\mathrm{Tr}\big(b\,\di (\bbvar{d} gg^{-1}) \big)\mp\sum_\pm\Big(p_A\bbvar{d}q^A-p_Aq^B \tensor{\big[g\,\bbvar{d}g^{-1}\big]}{^A_B}\Big)=\\
&= \mp\sum_\pm\Big(p_A\bbvar{d}q^A+\big(b_{AB}-q_{(A}p_{B)}\big)\big[g\,\bbvar{d}g^{-1}\big]^{AB}\Big).
\end{align}
The second term in the last line vanishes thanks to the gluing conditions \eref{glucond1} that provide junction conditions between the field theory in the interior and the boundary modes. Since the second term  vanishes, we are simply left with the boundary terms alone,
\begin{equation}
\Theta_\Sigma = \mp\sum_\pm p_A\bbvar{d}q^A.\label{thetadef}
\end{equation}
The only residual constraint is inherited from the gluing condition \eref{glucond1}, which translates now into the matching constraint
\begin{equation}
G_{AB}  = p^+_{(A}q^+_{B)} - p^-_{(A}q^-_{B)}=0.\label{Ccons}
\end{equation}
 This constraint generates simultaneous $SL(2,\R)$ transformations of the boundary modes $(p_A^\pm,q^A_\pm)$ on either end of the strip: from equation \eref{thetadef}, we infer the fundamental Poisson brackets
 \begin{equation}
 \big\{p^\pm_A,q_\pm^B\big\}=\mp\delta^B_A.
 \end{equation}
And for any $\ou{\lambda}{A}{B}\in\mathfrak{sl}(2,\R)$, we thus have
\begin{subalign}
\big\{G_{AB}\lambda^{AB},q^A_\pm\big\}&=+\ou{\lambda}{A}{B}q^B_\pm,\\
\big\{G_{AB}\lambda^{AB},p_A^\pm\big\}&=-\ou{\lambda}{B}{A}p_B^\pm.
\end{subalign}
The quantisation is straightforward. The kinematical Hilbert space is the tensor product $\mathcal{K}=L^2(\R^2,d^2q_+)\otimes L^2(\R^2,d^2q_-)$ with respect to the $SL(2,\R)$ invariant measure $d^2q=\tfrac{1}{2}\,\di q_A\wedge \di q^A$. On this Hilbert space, the matching constraint turns into the differential operator $\hat{G}_{AB}=\I\,q^+_{(A}\partial^+_{B)}-\I\,q^-_{(A}\partial^-_{B)}$ with $\partial^\pm_A$ denoting the partial derivatives  $\partial^\pm_A=\partial/\partial q^A_\pm$. Physical states are annihilated by the constraint and given by wave functions $\Psi_f(q_+,q_-)=f(q^+_Aq^A_-)$ that only depend on the $SL(2,\R)$ invariant contraction $x=\epsilon_{AB}q^A_+q^B_-$ of the configuration variables. The inner product between any two such  states is given by the integral
\begin{equation}
\langle\Psi_f,\Psi_{f'}\rangle = \int_{-\infty}^\infty \di x\,|x|\,\overline{f(x)}\,f^\prime(x),
\end{equation}
where the integration measure is found from the requirement that the $SL(2,\R)$ Casimir\footnote{Strictly speaking, the Casimir is the square of $p_Aq^A$.} $\widehat{p_A^\pm q^A_\pm}:\widehat{p_A^\pm q^A_\pm}\Psi_f=\pm\I (q^A_\pm\partial^\pm_A+1)\Psi_f=\pm\I(x\partial_x+1)f$ is self-adjoint.

\section{Quantisation and coupling to matter defects}\label{sec5}
\noindent At the end of the last section, we considered the quantisation of the bulk plus boundary modes in the absence of matter. The simplest way to include matter is to add point sources, which are charged under the gauge group. The coupled system is  governed by the action
\begin{align}\nonumber
S[B,A|P,Q,\psi] = \int_{\mathcal{M}}\ou{B}{A}{B}\ou{F}{B}{A}[A]-\int_\gamma\Big[\bar\psi_A D\psi^A-&\frac{N}{2}\big(\I \bar\psi_A\psi^A-s\big)\Big]+\\
&\mp\sum_\pm \int_{\gamma^\pm}\Big(P_ADQ^A -\kappa\,H_\pm[P,Q]\Big),\label{JTmatt}
\end{align} which is the sum of the gravitational bulk plus boundary action for JT gravity \eref{BFactn} and the action for the $\AdS$ particles \eref{partcl}, which are charged under $SL(2,\R)$. The manifold $\mathcal{M}$ contains  the worldine of the particle, which splits the manifold $\mathcal{M}$ into two adjacent regions $\mathcal{M}_+\cup\mathcal{M}_-=\mathcal{M}$, whose boundaries are $\partial\mathcal{M}_-=\gamma_-^{-1}\cup\gamma$ and $\partial\mathcal{M}_+=\gamma^{-1}\cup\gamma_+$ respectively.\footnote{N.B.\ we assume that the particle does not hit the boundary $\partial \mathcal{M} = \gamma_-^{-1}\cup\gamma_+$.}

The presence of a distributional matter current alters the field equations in the interior: it creates a discontinuity of the $B$ field, but does not affect the flatness constraint \eref{Flatnss}, which is unaffected by the matter coupling,\begin{align}
F_{AB} & = 0,\\
D_aB_{AB} & = -\delta_a[\gamma]\,\bar{\psi}_{(A}\psi_{B)}.\label{DBeq}
\end{align}
The one-form $\delta_a[\gamma]$ is a distribution,\footnote{The Dirac distribution $\tilde{\delta}(p,q)$ is a scalar density of weight one, and $\utilde{\epsilon}_{ab}$ denotes the inverse and two-dimensional Levi-Civita density, which is a tensor density of weight minus one.}
\begin{equation}
\delta_a[\gamma]\big|_p = \int_\gamma \di s\, \dot{\gamma}^b(s)\,\utilde{\epsilon}_{ab}\,\tilde{\delta}\big(p,\gamma(s)\big).
\end{equation}

Following the construction of the last section,  we introduce the phase space of the bulk plus boundary field theory. If $\Sigma$ denotes again a Cauchy hypersurface in $\Sigma$ that is anchored at the boundary $\gamma^{-1}_-\cup\gamma_+$, we find
\begin{equation}
\Theta_\Sigma = \int_\Sigma \ou{B}{A}{B}\bbvar{d}\ou{A}{A}{B}-\bar{\psi}_A\bbvar{d}\psi^A\big|_{\gamma\cap\Sigma}\mp\sum_\pm P_A\bbvar{d}Q^A\big|_{\gamma_\pm\cap\Sigma}.\label{symplctpot}
\end{equation}
To introduce coordinates on the physical phase space, we parametrise the flat $SL(2,\R)$ connection in terms of an $SL(2,\R)$ frame field $g:\mathcal{M}\rightarrow SL(2,\R)$ such that $A=g^{-1}\di g$. In the same way, we transform the $B$ field back into a constant Lie algebra element $b\in\mathfrak{sl}(2,\R)$, which  has a discontinuity across the worldine of the particle,
\begin{equation}
B = g^{-1}b_- g\,\Theta_-+g^{-1}b_+ g\,\Theta_-,\quad \di b_\pm=0,\label{Bparam}
\end{equation}
with $\Theta_\pm$ denoting the step function,
\begin{equation}
\Theta_\pm(p)=\begin{cases}
+1,\,\text{if}\; p\in\mathcal{M}_\pm,\\
0,\,\text{otherwise}.
\end{cases}
\end{equation}
The strength of the discontinuity is determined by the matter current: going back to \eref{DBeq}, and integrating $\di b$ along various intervals in $\Sigma$, we obtain the junction conditions\begin{subalign}
G_{AB}&=p^{+}_{(A}q^+_{B)}-p^{-}_{(A}q^-_{B)}+\bar{\varphi}_{(A}\varphi_{B)}=0,\label{match1}\\
b^\pm_{AB}&=p^{\pm}_{(A}q^\pm_{B)},\label{match2}
\end{subalign}
where we introduced the dressed boundary modes
\begin{subalign}
(p_A, q^A) & = (\ou{[g^{-1}]}{B}{A}P_B,\ou{g}{A}{B}Q^B),\label{dressdvar1}\\
\varphi^A &:= \ou{g}{A}{B}\psi^B.\label{dressd3}
\end{subalign}
The physical phase space is obtained then by imposing the matching constraint \eref{match1} at the level of the kinematical phase space, which is equipped with a natural symplectic structure: if we express the pre-symplectic potential in terms of the dressed boundary modes (\ref{dressdvar1}, \ref{dressd3}), we find
\begin{equation}
\Theta_\Sigma = -\bar{\varphi}_A\bbvar{d}\varphi^A\big|_{\gamma\cap\Sigma}\mp\sum_\pm p_A\bbvar{d}q^A\big|_{\gamma_\pm\cap\Sigma},\label{thpotntl}
\end{equation}
which generalises the previous result \eref{thetadef} to include distributional matter sources.

In the following, we restrict ourselves to the region in phase space where $x=\epsilon_{AB}q_+^Aq_-^B>0$. Since the $\R^2$ spinors $q^A_\pm$ are now linearly independent, they form a basis such that we can expand $\varphi^A$ in terms of the JT gravitational edge modes $q^A_\pm$,\begin{equation}
\varphi^A = \frac{1}{\sqrt{x}}\big( zq^A_+ +\bar{p} q^A_-),\label{param1}
\end{equation}
where $\bar{p}\in\C$ and $z\in\C$ are the spin up and down components of $\varphi^A$ with respect to the basis $(q^A_+,q^A_-)$. Next, we turn to the symplectic structure of the matter sector, and write it in terms of the new canonical pair $(p,z)$ and the JT boundary modes $q^A_\pm$. Inserting \eref{param1} into $\bar{\varphi}_A\bbvar{d}\varphi^A$, we obtain
\begin{align*}
\bar{\varphi}_A\bbvar{d}\varphi^A & = -p\bbvar{d} z +\bar{z}\bbvar{d} \bar{p} + \frac{1}{2x}(pz-\bar{p}\bar{z})\bbvar{d}{x}+\\
&\hspace{4em}+\frac{z}{x}\big(\bar{z}q^+_A+pq^-_A\big)\bbvar{d}q^A_++\frac{\bar{p}}{x}\big(\bar{z}q^+_A+pq^-_A\big)\bbvar{d}q^A_-=\\
& = -p\bbvar{d} z +\bar{z}\bbvar{d} \bar{p}+ \frac{1}{x}\Big(z\bar{z}\,q^+_A+\frac{1}{2}(pz+\bar{p}\bar{z})q^-_A\Big)\bbvar{d}q^A_+
+\frac{1}{x}\Big(p\bar{p}\,q^-_A+\frac{1}{2}(pz+\bar{p}\bar{z})q^+_A\Big)\bbvar{d}q^A_-.
\end{align*}
Next, we introduce the shifted momentum variables
\begin{subalign}
\pi^+_A & = p^+_A+ \frac{1}{x}\Big(z\bar{z}\,q^+_A+\frac{1}{2}(pz+\bar{p}\bar{z})q^-_A\Big),\\
\pi^-_A & = p^-_A- \frac{1}{x}\Big(p\bar{p}\,q^-_A+\frac{1}{2}(pz+\bar{p}\bar{z})q^+_A\Big).
\end{subalign}
The new variables allow us to bring the symplectic potential \eref{thpotntl} into the following compact form,
\begin{equation}
\Theta_\Sigma = p\bbvar{d} z -\bar{z}\bbvar{d} \bar{p}-\pi^+_A\bbvar{d} q_+^A+\pi^-_A\bbvar{d} q_-^A.
\end{equation}
Therefore, the map from $(\bar{\varphi}_A,\varphi^A,p_A^\pm,q^A_\pm)$ to $(p,\bar{p},z,\bar{z},\pi_A^\pm,q^A_\pm)$ is a canonical transformation. In terms of the new canonical variables, the junction condition is now simply given by
\begin{equation}
G_{AB} = \pi^+_{(A}q^+_{B)}-\pi^-_{(A}q^-_{B)},\label{constrnd}
\end{equation}
such that we have eliminated the $\bar{\varphi}_{(A}\varphi_{B)}$ term from the constraint \eref{match1}. 

As in the vacuum case, the quantisation is straightforward. Consider a Schrödinger position representation. Kinematical states are now vectors in the auxiliary Hilbert space $\mathcal{K}$, which is the tensor product $\mathcal{K}=L^2(\R^2,d^2q_+)\otimes L^2(\R^2,d^2q_-)\otimes L^2(\C,\tfrac{\I}{2}\di z\wedge\di\bar{z})$. On this Hilbert space, the configuration variables $(q^A_\pm,z,\bar{z})$ act by multiplication, the conjugate momenta by taking the derivative, such that
\begin{align}
\hat{\pi}^\pm_A &= \pm\I\frac{\partial}{\partial q^A_\pm} =: \pm\I\partial^\pm_A,\\
\hat{p}&=-\I\partial_{{z}},\quad \hat{\bar{p}} = -\I\partial_{\bar{z}}.
\end{align}
The junction conditions are now operator-valued and given by the $SL(2,\R)$ generators
\begin{equation}
\hat{G}_{AB}=+\I q^+_{(A}\partial^+_{B)}+\I q^-_{(A}\partial^-_{B)}.
\end{equation}
Physical states $\Psi(q^A_+,q^A_-,z,\bar{z})$ lie in the kernel of $\hat{G}$ and must be therefore functions of $z$, $\bar{z}$ and the $SL(2,\R)$ invariant contraction $x=q_A^+q^A_-$ alone,
\begin{equation}
\Psi_f(q^A_+,q^A_-,z) = f(q_A^+q^A_-,z).\label{innprod}
\end{equation}
We have restricted ourselves to a region of phase space, where $x=q_A^+q^A_->0$ and the inner product is therefore given by\footnote{If we include also the region in phase space, where $x<0$, we would end up with two-component wave functions, where the up (down) components correspond to the regions $x> 0$ ($x<0$) of phase space respectively.}
\begin{equation}
\langle\Psi_{f'},\Psi_f \rangle = \frac{\I}{2}\int_\C\di z\,\di\bar{z}\int_0^\infty \di x\, x\,\overline{f'(x,z)}f(x,z).
\end{equation}
An orthonormal basis in the physical Hilbert space can be constructed as follows. Consider first the $SL(2,\R)$ Casimir on either end of the boundary,
\begin{equation}
\rho_\pm=p^\pm_Aq_\pm^A = \pi^\pm_Aq_\pm^A +\frac{1}{2}(pz+\bar{p}\bar{z}).
\end{equation}
In terms of the Schrödinger representation, the $SL(2,\R)$ Casimir $\rho_\pm$ is a sum of dilation operators,
\begin{equation}
\hat{\rho}_\pm=\pm\I(q^A_\pm\partial^\pm_A+1)+\frac{1}{2\I}(z\partial_z+\bar{z}\partial_{\bar{z}}+1)=\pm\I(x\partial_x+1)+\frac{1}{2\I}(z\partial_z+\bar{z}\partial_{\bar{z}}+1),
\end{equation}
where we chose a symmetric ordering for the operator product $\widehat{q^A_\pm p_A^\pm}$. In the same way, we consider now the Casimir for the discrete series representation. Going back to the parametrisation \eref{param1} of $\varphi^A$ in terms of $q^A_\pm$, we find
\begin{equation}
s= \I\bar{\psi}_A\psi^A = -\I(pz-\bar{p}\bar{z}).
\end{equation}
In quantum theory, the imaginary part of $pz$ is nothing but the angular momentum operator on the complex plane,
\begin{equation}
\hat{s}=-(z\partial_z-\bar{z}\partial_{\bar{z}}),
\end{equation}
whose spectrum is discrete $s\in\Z$. An orthonormal basis is now given by the homogenous functions 
\begin{equation}
\big\langle q_+^A,q_-^A,z,\bar{z}\big|\rho_+,\rho_-,s\big\rangle  = 
\frac{1}{(2\pi)^{\frac{3}{2}}} \Theta(\epsilon_{AB}q^A_+q^B_-)\,\big(\epsilon_{AB}q^A_+q^B_-\big)^{-\I\varepsilon-1}\,(z\bar{z})^{\I\rho-\frac{1}{2}}\left(\frac{\bar{z}}{z}\right)^{\frac{s}{2}},\label{physstate}
\end{equation}
where
\begin{equation}
\rho_\pm = \rho\pm\varepsilon.\label{massjump}
\end{equation}
Given the inner product \eref{innprod}, the states are normalised as
\begin{equation}
\big\langle \rho_+,\rho_-,s\big|\rho_+',\rho_-',s'\big\rangle =\delta_{ss'}\delta(\rho_+-\rho_+')\delta(\rho_--\rho_-').
\end{equation}
The jump \eref{massjump} is a jump in energy: the Hamiltonian $\boldsymbol{H}_\xi$  is the sum of boundary terms, $\boldsymbol{H}_\xi=\mp\sum_\pm\kappa_\xi f_\pm(P_AQ^A)\big|_{\gamma_\pm\cup\Sigma}$. A simple choice for the boundary Hamiltonian, which is often used in the literature \cite{Cangemi:1992bj,Iliesiu:2019xuh,Maldacena:2016upp}, is to identify the Hamiltonian with the $SL(2,\R)$ Casimir. In other words $f_\pm(P_AQ^A)= (P_AQ^A)^2$, such that
\begin{equation}
{\boldsymbol{\hat{H}}}_\xi = \kappa_\xi^-\big(\hat{\rho}_-\big)^2-\kappa_\xi^+\big(\hat{\rho}_+\big)^2.
\end{equation}

\section{Summary and Conclusion}
\noindent Let us  summarise. In this paper, we developed a twistor quantisation of JT gravity with distributional matter defects. 
 At the boundary of the manifold, which has the topology of a strip $[-1,1]\times\R$, we chose specific Dirichlet boundary conditions, where the connection at the boundary is unconstrained.  Such $SL(2,\R)$ gauge covariant boundary conditions are possible only at the expense of working on an extended phase space with additional boundary degrees of freedom  \cite{Donnelly:2016auv,Wieland:2017zkf}. We then saw that these boundary modes can be encoded into $SL(2,\R)$ boundary spinors $(P_A,Q^A)\in (\R^2)^\ast\oplus\R^2$, which are the eigen vectors of the $\mathfrak{sl}(2,\R)$ valued $B$ field. Using these boundary fields, we introduced the boundary term that makes the variational problem well posed. The boundary term consists of a kinetic term $P_ADQ^A$ plus a Hamiltonian $H[P,Q]$, which characterises the quasi-local energy of the  bulk plus boundary system. After completing the Hamiltonian analysis, we considered the system at the quantum level. Physical states are constructed by fusing irreducible unitary representation of $SL(2,\R)$ into gauge invariant singlets. The continuous (principal) series representations of $SL(2,\R)$ describe empty patches of $\AdS$. Spinning point particles, on the other hand, are characterised by the discrete series representations. Along the worldlines of such particles, the  $B$ field has a discontinuity. The strength of the discontinuity is governed by the junction conditions \eref{match1} that fuse the three $SL(2,\R)$ representations into an $SL(2,\R)$ invariant singlet \eref{physstate}. 

Our quantisation  is based on a relatively simple Schrö\-dinger representation. The kinematical Hilbert space for every patch of the boundary is  $L^2(\R^2,d^2q_-)$ $\otimes L^2(\R^2,d^2q_+)$, where $q^A_\pm$ are the JT boundary modes at the ends of the interval. The canonical momentum variables $p_A^\pm=\pm\I\partial/\partial q^A_\pm$ are commuting. In  comparison, the more familiar $SL(2,\R)$ holonomy representation \cite{Iliesiu:2019xuh,JTdefects} is based on square integrable functions  $\Psi\in L^2(SL(2,\R),d\mu)$.\footnote{Since the group is non-compact, the invariant measure is unique only up to an overall constant,  $d\mu \propto \mathrm{Tr}(g^{-1}\di g\wedge g^{-1}\di g\wedge g^{-1}\di g)$.} The corresponding classical phase space  is $T^\ast SL(2,\R)$ and the momentum variables, which are left invariant vector fields, are non-commutative. The situation is reminiscent of what we know from loop quantum gravity. The holonomy representation \cite{Iliesiu:2019xuh,JTdefects} of JT gravity is the analogue of the spin network representation. The twistor representation of JT gravity, on the other hand, is analogous to the spinor representation of spin network states \cite{twist,Livinerep,Bianchi:2016hmk,Wieland:2017cmf,komplexspinors}, where every gravitational Wilson line splits into a pair of surface charges, which are entangled across the connecting link \cite{twist,Bianchi:2016hmk}. In the bulk, such surface charges are not directly observable. At a physical boundary, which could be an isolated horizon or any other entangling surface, they are. In fact, operators for quasi-local energy and angular momentum  are quadratic invariants of the boundary spinors \cite{Wieland:2017zkf}. The energy for a stationary observer in the vicinity of an isolated horizon is nothing but the area \cite{FGPfirstlaw}, which turns into a dilation operator upon quantization \cite{Wieland:2017cmf}. The resulting eigen functions  are homogenous functions of the boundary modes and they carry a unitary irreducible representation of $SL(2,\C)$. In JT gravity we found a very similar result. Eigenfunctions of quasi-local energy are homogenous functions of the spinor-valued boundary modes \eref{physstate} and they carry a unitary  representation of $SL(2,\R)$. 

In this paper, we constructed a twistor representation for JT gravity coupled to point particles.\footnote{We considered only a single such point source, but the generalisation to an arbitrary number of particles is straightforward.} We have not studied, however, the corresponding scattering amplitudes in the JT twistor approach. This does not seem too difficult a problem, because the amplitudes must be $SL(2,\R)$  invariant functionals of the boundary data. Such functionals can be easily classified in terms of the $SL(2,\R)$ invariant cross ratios $x_{ij}=\epsilon_{AB}q^A_iq^B_j$ of the boundary data $q^A_i$. Spinfoam amplitues \cite{alexreview} and the more ordinary path integral approaches to JT gravity \cite{Mertens:2017mtv,Blommaert:2018oro,Lam:2018pvp} are based on very similar techniques. It would be interesting to explore this connection in more detail. 


\paragraph{Acknowledgments} I thank Laurent Freidel, and William Donnelly 
for many encouraging discussions on the topic of this paper. Part of this research was completed during my visit to the University of Graz in early March 2020. I would like to thank Reinhard Alkofer for support during this period. This research was supported by Perimeter Institute. Research at Perimeter Institute is supported in part by the Government of Canada through the Department of Innovation, Science and Economic Development Canada and by the Province of Ontario through the Ministry of Colleges and Universities.
\providecommand{\href}[2]{#2}\begingroup\raggedright\endgroup

\end{document}